\documentclass[letterpaper,twocolumn,10pt]{article}
\usepackage{usenix}
\usepackage{tikz}
\usepackage{amsmath}

\usepackage{fancyhdr}
\usepackage{url}
\usepackage{enumitem} 
\usepackage{subcaption}
\usepackage{multirow}

\usepackage{hyperref}
\usepackage{algorithm}
\usepackage {amssymb}
\usepackage{graphicx}
\usepackage{algorithmic}
\usepackage{balance}
\usepackage{listings}
\usepackage{xcolor}
\usepackage{lineno}
\usepackage[normalem]{ulem}
\usepackage{url}
\usepackage{breakurl}
\usepackage{hyperref}
\usepackage{titling}
\usepackage{setspace}
\usepackage{wrapfig}
\usepackage{bbding}
\usepackage{booktabs}
\usepackage{xspace}
\usepackage{makecell}
\usepackage{pifont} 
\usepackage{tikz}
\usepackage{xcolor}

\newcommand{\circleNum}[2][1.2em]{
	\tikz[baseline=(char.base)]{
		\node[shape=circle, fill=black, text=white, inner sep=0pt,
		minimum size=#1, font=\bfseries] (char) {#2};
	}
}

\newcommand{\circleOne}[2][1.2em]{
	\tikz[baseline=(char.base)]{
		\node[shape=circle, draw, inner sep=0pt,
		minimum size=#1] (char) {#2};
	}
}

\makeatletter
\renewcommand{\fnum@figure}{\textbf{\figurename~\thefigure}}
\renewcommand{\fnum@table}{\textbf{\tablename~\thetable}}
\makeatother
\newcommand{\sysname}{\textsc{Helmsman}\xspace}

\begin{document}

\date{}

\title{\Large \bf The Clustering Strikes Back:\\
Building Cost-Effective and High-Performance ANNS at Scale with Helmsman
}
	
\author{
	Yuchen Huang\textsuperscript{1,3},
	Baiteng Ma\textsuperscript{1,3},
	Yiping Sun\textsuperscript{3},
	Yang Shi\textsuperscript{3},
	Xiao Chen\textsuperscript{3},\\
	Xiaocheng Zhong\textsuperscript{3},
	Zhiyong Wang\textsuperscript{3},
	Yao Hu\textsuperscript{3},
	Erci Xu\textsuperscript{2},
	Chuliang Weng\textsuperscript{1}\\
	\textsuperscript{1}East China Normal University
	\qquad
	\textsuperscript{2}Shanghai Jiaotong University
	\qquad
	\textsuperscript{3}Xiaohongshu Inc\\
}

\maketitle

\begin{abstract}
RedNote (a.k.a., Xiaohongshu, a global-scale social network platform) widely
adopts approximate nearest neighbor search (ANNS) to power its search,
recommendation, and advertising services. Due to the demanding Service Level
Agreements (SLAs), we have to rely on in-memory graph-based ANNS (i.e., HNSW)
to provide high throughput and low latency.

However, the ever-growing user base and content volume have led to an explosive
increase in memory footprint and consequently huge CapEx and OpEx. After
exploring various alternatives, we find that building a clustering-based ANNS on
top of all-flash servers can be promising. Yet, we still experience severe
overheads from the kernel I/O stack, a fixed pruning strategy, and slow index
construction.

We present \sysname, a high-performance and cost-effective clustering-based ANNS
system, which combines an ANNS-oriented userspace storage stack, a
leveling-learned pruning module, and GPU-accelerated pipelines of construction.
\sysname saves over 90\% of hardware costs and enables billion-scale index
(re)builds within hours. 
In the current production deployment, operating
stably for several months, 40 machines now host ANNS workloads that previously
required about 35{,}000 cores and 0.35 PB DRAM.
\end{abstract}
\thispagestyle{empty}

\section{Introduction}
RedNote (a.k.a., Xiaohongshu)~\cite{RedNote} is a global social media platform
for users to share and interact with photos, videos, and text in the form of
``notes''. By 2025, the platform hosts over 300 million monthly active users and
more than 80 million creators~\cite{restofworld-xiaohongshu-2025}. As the core
infrastructure behind search, recommendation, and
advertising~\cite{ParlayANN,Quake,PipeANN}, our approximate nearest neighbor
search (ANNS) service manages hundreds of billions of embedding vectors and
sustains millions of queries per second (QPS) with strict latency service-level
agreements (SLAs) (e.g., 5-10 ms for real-time services).

To meet the quality of services (QoS), we previously relied on large in-DRAM
graph indexes~\cite{HNSW,NSG} for high throughput and low latency.  However, the
scaling of users and content drastically increases the memory footprint, pushing
the ANNS infrastructure at RedNote to reach the petabyte level. This incurs
prohibitively high CapEx and OpEx for the existing in-DRAM ANNS infrastructure~\cite{kioxia-vectordb-2024,opensearch-disk-ann-2025}.  We are
therefore motivated to explore more cost-efficient alternatives. 

The recent advancement in NVMe SSDs has made them a promising candidate given
their high bandwidth and low cost per GB~\cite{NVMeArrays,ModernNVMe,StructuredStorage}.  For
example,  a 12-drive PCIe-Gen5 SSD array~\cite{PM9D3A} can deliver roughly 30\%
of the bandwidth of a 12-channel DDR5 memory~\cite{DDR5RDIMM}, yet SSD costs
about 1/40 of DRAM (i.e., 0.2 \$/GB vs. 8 \$/GB). While we have successfully
deployed graph-based hybrid (SSD+DRAM) ANNS systems like DiskANN~\cite{DiskANN}
for the more relaxed offline workloads (e.g., content moderation), using them for online
services (e.g., search) remains impracticable in terms of
latency and throughput SLAs. The main reason is the greedy graph traversals
maintain long candidate lists (e.g., lengths can be up to 1.5$\times$ top-$k$),
which incur many serialized I/Os and thus fail to utilize high bandwidth of
SSDs.

The above observation motivates us to rethink the possibility of using
clustering-based ANNS.  With dependency-free I/Os issued in batches,
clustering-based ANNS, such as SPANN~\cite{SPANN}, can be expected to achieve
low latency that meets our SLA and also be scaled for throughput by adding
more SSDs.  Yet, to apply them in production remains challenging.  First, under
the traditional I/O stack, SPANN uses only about 20–60\% of SSD bandwidth,
resulting in a significant throughput gap to in-DRAM HNSW.  Second, given the varying
top-$k$ of real workloads, the existing fixed pruning approach cannot
sufficiently reduce redundant scans and leads to unstable per-query performance.
Third, the existing single-node and CPU-only construction can take tens of hours
or even days, failing to support fast rebuilds needed by frequent updates of
embedding models and vectors.

\begin{figure*}[t]
	\centering
	\begin{subfigure}[t]{0.24\linewidth}
		\includegraphics[width=\linewidth]{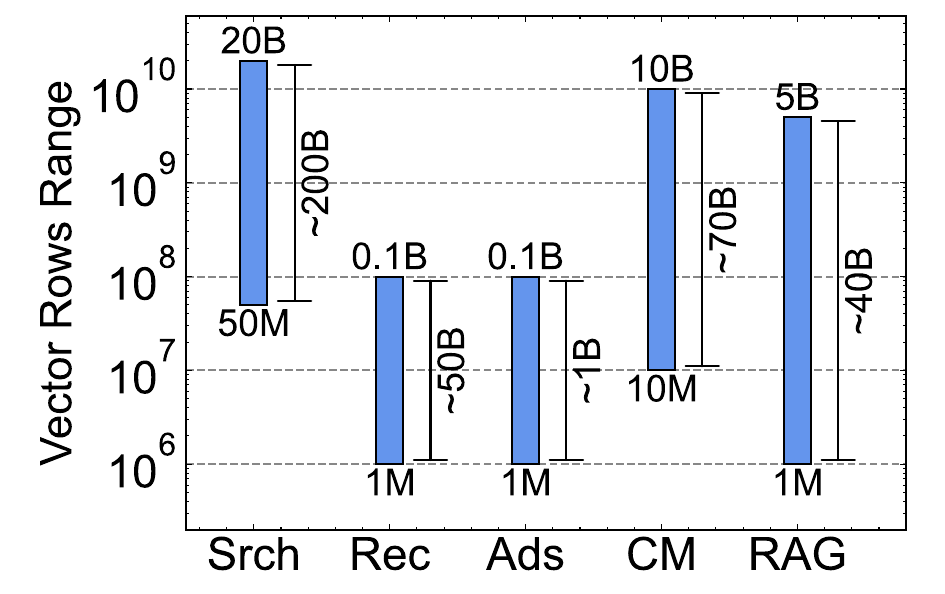}
		\caption{\textit{{Total rows of vector datasets.}}}
		\label{fig:bak1}

	\end{subfigure}
	\centering
	\begin{subfigure}[t]{0.24\linewidth}
		\includegraphics[width=\linewidth]{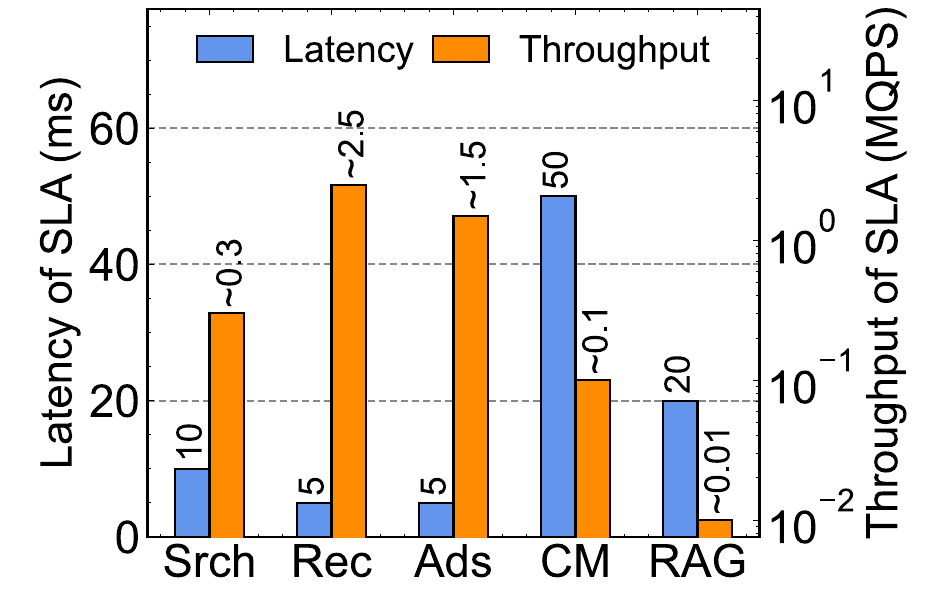}
		\caption{\textit{{SLAs of performance.}}}
		\label{fig:bak2}

	\end{subfigure}
	\begin{subfigure}[t]{0.24\linewidth}
		\includegraphics[width=\linewidth]{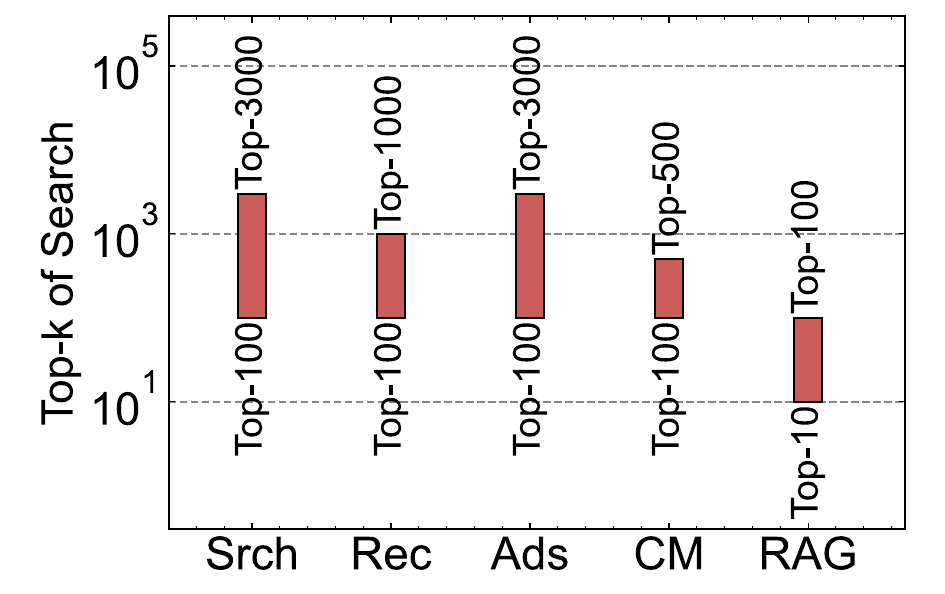}
		\caption{\textit{{Top-k value of requests.}}}
		\label{fig:bak3}

	\end{subfigure}
	\begin{subfigure}[t]{0.24\linewidth}
		\includegraphics[width=\linewidth]{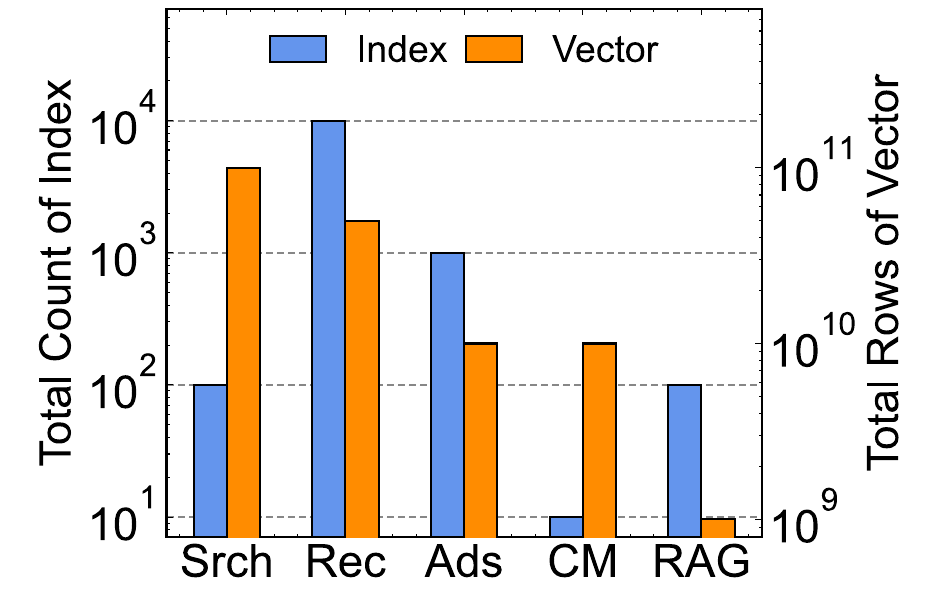}
		\caption{\textit{{Frequency of construction.}}}
		\label{fig:bak3-2}

	\end{subfigure}
	\centering
	\caption{\textit{Characterizing ANNS of RedNote along 4 aspects: scale, SLA,
			top-k, and construction. Search is denoted as \textsf{Srch}, and others are
			recommendation (\textsf{Rec}), advertising (\textsf{Ads}), content moderation
			(\textsf{CM}), and retrieval-augmented generation (\textsf{RAG}).}}
	\label{fig:bak-over}
	\vspace{-3mm}
\end{figure*}

Hence, to overcome these limitations, we develop \sysname, a high-performance and
cost-effective ANNS system using clustering-based indexes atop all-flash
servers. There are three main techniques.  First, we build an ANNS-oriented
userspace storage backend that bypasses the kernel I/O stack, minimizing
software overhead and directly orchestrating devices to match the ANNS-specific
I/O patterns.  Second, we develop a leveling-learned pruning module that adapts
to both top-$k$ and query distributions, while remaining compatible with
SSD-friendly batched I/O. Third, we leverage GPU acceleration and dynamically
allocate CPUs for index construction, enabling sub-hour builds for
common-scale (e.g., 0.1B) and hour-level builds for ultra-large-scale (e.g., 10B) indexes.

Our evaluations show that \sysname delivers 2–16$\times$ higher throughput than
existing DRAM–SSD ANNS and up to 85\% of the throughput of in-DRAM ANNS, while
preserving latency SLAs. We are actively rolling out \sysname to our production
environment. Currently, \sysname can use only 40 machines with about 30-40 TB DRAM in total
to sustain online traffic that previously consumed roughly 35{,}000 CPU cores
and $\sim$0.35 PB of in-DRAM ANNS infrastructure, saving more than 90\% device
costs.
The open-source proof-of-concept version of \sysname and datasets fitted to real-world distributions are available at {\url{https://github.com/Red-EAD/helmsman}}.

\section{Background}
\label{sec:background}

\subsection{ANNS-based Services in RedNote}
\label{subsec:xhsanns}
RedNote (a.k.a., Xiaohongshu)~\cite{RedNote} is a global social platform with
hundreds of millions of active users sharing and interacting with various
types of content including pictures, comments and short videos on a daily basis.
The proper functioning of RedNote relies heavily on the search engine,
recommendation system, advertising platform, content moderation, and emerging AI
services (e.g., retrieval-augmented generation for large language models). All
of the above are supported by approximate nearest neighbor search (ANNS) over
hundreds of billions of high-dimensional vectors with millions of queries per
second.  Despite all being ANNS-based, these services can have different
characteristics as shown in~\autoref{fig:bak-over}, such as scale of dataset
(\autoref{fig:bak1}), the SLA of performance (\autoref{fig:bak2}), the top-$k$
of requests (\autoref{fig:bak3}), and the index building (\autoref{fig:bak3-2}).
We now further discuss them in detail.

\noindent\textbf{Search.} First, our search targets the full corpus (i.e., all
user-generated content and web-wide data) and the index can reach up to tens of
billions of vectors. In addition, the search services also follow a multi-path
workflow where different recall paths cover various subsets of the corpus (e.g.,
texts, images, videos, and user behaviors) to ensure high accuracy.  Therefore,
search has the largest overall vector volume, totaling up to twenty billion
vectors in practice. Second, search is directly exposed to end users, which
requires an SLA of $\sim$10 millisecond average latency~\cite{SrchLa} under
$\sim$300K QPS, our typical peak traffic. Third, ANNS is the first stage of the
ranking pipeline in search. Thus, it requires a relatively large top-$k$
(e.g., around $100$ to $3{,}000$) to provide sufficient candidates for
downstream re-ranking models~\cite{TaobaoEBR,CLEAR}. Finally, because our
embedding models are continuously trained with daily collected metrics (e.g.,
user behaviors and content popularity), we wish to rebuild (instead of in-place
updating such as SPFresh~\cite{SPFresh}, Quake~\cite{Quake} and
OdinANN~\cite{OdinANN}) the entire indexes on a daily basis.

\noindent\textbf{Recommendation.} Unlike search, recommendation only focuses on
the popular subsets of full corpus. As a result, the total vector volume is
significantly smaller than that of search, ranging from 1 to 100 million vectors per
index. Recommendation directly serves end users but involves more recall paths.
Thus, the online request throughput can reach ~$\sim$2.5 million QPS while still
requiring millisecond-level latency SLAs~\cite{RecLa}. Moreover, post-filtering
based on user-specific constraints is common in recommendation.  To ensure
sufficient results after filtering~\cite{VBASE}, the top-$k$ from ANNS can be up
to $1{,}000$. Finally, embedding models are updated in batches (aggregated from
minutes to hours~\cite{QuickUpdate,Ekko}) based on real-time feedback (e.g.,
click-through rates), leading to up to ten thousand index rebuilds per day and a
cumulative rebuilt volume on the order of tens of billions of vectors.

\noindent\textbf{Advertising.} Embedded from products, advertising has fewer
vectors, around one billion in total. Similar to search and recommendation,
advertising also directly serves users and thus requires millisecond-level
latency SLAs under millions of QPS~\cite{AdsLa}. In addition, advertising 
follows a post-filtering pipeline. ANNS retrieval is followed by scalar
filtering on multiple attributes (e.g., exposure constraints and billing
status), so the required top-$k$ can be large as well (e.g., up to
3,000)~\cite{SBCN}. Finally, for index construction, since the embedding models
are modified incrementally by rates of clicks or purchases on the order of
minutes, this requires that index rebuilds keep up with these real-time updates as
well.

\noindent\textbf{Content moderation}. This maintains two types of indexes, the
large allow-list corpus and the smaller block-list corpus. The allowlists
consist of all content conforming to policies and values, so their indexes can
reach up to 10B vectors. The block-list stores only restricted content and is
typically much smaller (e.g., 10M vectors). Content moderation is mainly an
offline workload, and new content is compared against both allow-list and
block-list, then assessment models select high-risk candidates that are
forwarded to manual checking. This pipeline results in more relaxed latency SLAs
and lower overall throughput than the above-mentioned services~\cite{CMLa}.
Additionally, to capture restricted content with best effort, we also use
relatively large top-$k$ (up to $\sim$500). Since these collections change
slowly, block-list indexes are rebuilt on a daily basis and the allow-list
indexes are refreshed weekly.

\noindent\textbf{Retrieval-augmented generation}. Our ANNS also supports
emerging AI applications like RAG. In RAG, ANNS is mainly used to index
task-specific knowledge bases with sizes varying widely from about one million
to several billion vectors depending on the applications.  Since LLM inference
dominates resource consumption and current models have limited context length,
the SLA requirements for the ANNS stage are relatively relaxed (e.g., $\sim$20
ms)~\cite{RagLa} and the required top-$k$ is much smaller (e.g., 10-100) than
other services'~\cite{KGFiD,CoRAG}.  Moreover, the underlying knowledge bases
evolve slowly and the embedding models change infrequently, so index rebuilds
are needed far less often than in other online workloads.

\begin{figure}[t]
	\centering
	\begin{subfigure}[t]{0.48\linewidth}
		\includegraphics[width=\linewidth]{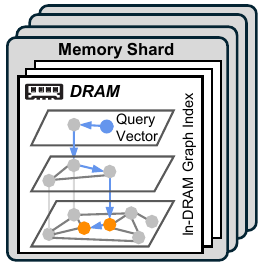}
		\caption{\textit{{In-DRAM graph indexes on distributed nodes for
					achieving low latency.}}}
		\label{fig:bak4}
	\end{subfigure}
	\centering
	\begin{subfigure}[t]{0.48\linewidth}
		\includegraphics[width=\linewidth]{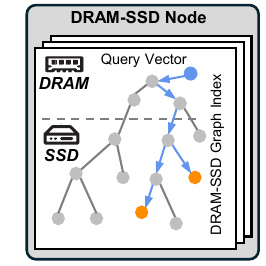}
		\caption{\textit{{DRAM-SSD graph indexes on a single node for saving
					memory footprint.}}}
		\label{fig:bak5}
	\end{subfigure}
	\centering
	\caption{\textit{Two solutions employed in services of RedNote.}}
	\label{fig:bak}
		\vspace{-5mm}
\end{figure}

\subsection{Solutions for ANNS-based Services}
Graph-based ANNS offers high throughput with low latency, and is therefore
widely deployed in RedNote’s production services. To satisfy the strict
performance SLAs of search, recommendation, and advertising, we run a large
fleet of distributed memory shards that host in-DRAM graph
indexes~\cite{HNSW,NSG}. The aggregated count of CPU cores already exceeds
10$^5$ among $4{,}000$ nodes, spanning $\sim$50 clusters as of today.  For
scenarios with slightly more relaxed latency and throughput requirements (e.g.,
content moderation and RAG), we instead employ hybrid DRAM–SSD nodes
with SSD-backed graph indexes~\cite{DiskANN,Starling,PipeANN} to trade some
latency for a much smaller memory footprint.

\noindent\textbf{In-DRAM Graph.} In our production, HNSW~\cite{HNSW}, one of the
state-of-the-art in-DRAM indexes, is commonly adopted.  And it is widely
deployed for search, recommendation, and advertising. \autoref{fig:bak4} shows,
for a given query, the HNSW search starts from an entry node on the top layer
and performs greedy best-neighbor descent layer by layer, until it reaches the
bottom, where a best-first search over a candidate pool produces the final
top-$k$ results.  All graph data, including neighbor lists and raw vectors, are
kept in DRAM so that the critical path consists only of in-memory accesses. When
a single shard cannot hold the entire index, we partition the dataset across
multiple memory shards and execute search independently on each HNSW shard. The
frontend then merges the partial results from all shards to obtain the global
top-$k$ answers. This design can deliver consistently low search latency and
high QPS, guaranteeing QoS.

\noindent\textbf{Hybrid DRAM-SSD Graph.} For scenarios with more relaxed
performance SLAs, such as content moderation and RAG, we employ hybrid
DRAM–SSD graph indexes, as illustrated in ~\autoref{fig:bak5}. These indexes
significantly reduce DRAM consumption, allowing a large-scale graph to be hosted
on a single node. Representative designs include DiskANN~\cite{DiskANN} and its
recent optimized variants such as Starling~\cite{Starling} and
PipeANN~\cite{PipeANN}. They retain PQ-compressed vectors and a small set of
frequently accessed vectors cached in DRAM, while storing the full-precision
vectors and graph edges on SSDs to cut memory footprint. During query
processing, the graph is traversed with the best-first search strategy and a
beam-search procedure. The search path always advances along the edges towards
the neighbor that is closest to the query. The beam-search width is translated
into an I/O width that controls how many candidates are fetched from SSDs in
batches, thereby improving bandwidth utilization for higher performance.

\section{Motivation}
\begin{figure}[t]
	\centering
	\begin{subfigure}[t]{0.48\linewidth}
		\includegraphics[width=\linewidth]{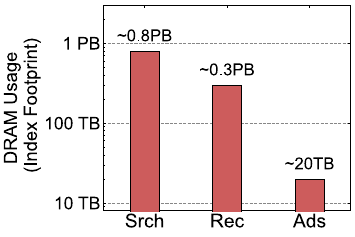}
		\caption{\textit{{PB-level DRAM usage.}}}
		\label{fig:mot1}
	\end{subfigure}
	\centering
	\begin{subfigure}[t]{0.48\linewidth}
		\includegraphics[width=\linewidth]{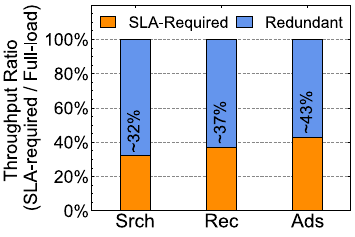}
		\caption{\textit{Throughput overkill.}}
		\label{fig:mot2}
	\end{subfigure}
	\centering
	\caption{\textit{In our online services, latency SLAs force massive DRAM for
	pure in-DRAM indexes, while throughput SLAs are modest to the full-load
	maximum of the in-DRAM deployment.}}
	\label{fig:mot}
		\vspace{-5mm}
\end{figure}
\subsection{Can't Afford In-memory ANNS to Scale}
Over the past few years, the fast growth of RedNote's user base has driven the
data volume of our stored vector corpus to be doubled annually.  The more than
billions of new vectors persisted each year exacts a heavy toll on the CapEx of
our in-memory ANNS deployment. \autoref{fig:mot1} illustrates that, as of 2025,
the in-DRAM HNSW indexes for search, recommendation, and advertising have
already consumed PB-level DRAM in practice.  Even just the ANN indexes of a
single service (e.g., search) can now consume hundreds of terabytes of DRAM.

However, the massive DRAM usage is mostly provisioned to maintain the QoS (e.g.,
10 ms average latency SLA of search). Specifically, \autoref{fig:mot2} shows
that, out of the 100\% throughput provided by in-DRAM deployments, only
$\sim$32–43\% is actually needed to maintain the required latency SLAs.
Nevertheless, we still need the rest 57–68\% of DRAM, not for throughput, but
for the capacity to host the entire set of indexes to avoid latency spikes under
workload bursts~\cite{HVS}.

\subsection{Can't Replace with Hybrid Graph}
\begin{figure}[t]
	\centering
	\begin{subfigure}[t]{0.32\linewidth}
			\includegraphics[width=\linewidth]{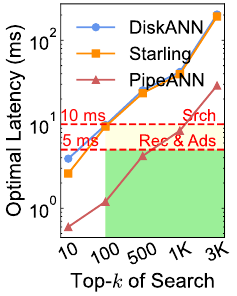}
			\caption{\textit{Average latency.}}
			\label{fig:mot4-1}
		\end{subfigure}
	\centering
\begin{subfigure}[t]{0.32\linewidth}
	\includegraphics[width=\linewidth]{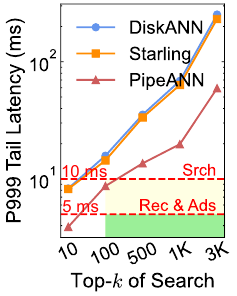}
	\caption{\textit{{P999 latency.}}}
	\label{fig:mot4-2}
\end{subfigure}
	\centering
	\begin{subfigure}[t]{0.32\linewidth}
			\includegraphics[width=\linewidth]{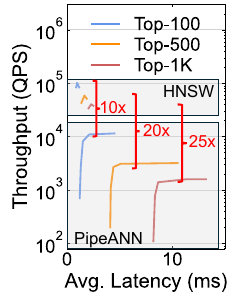}
			\caption{\textit{Throughput gap.}}
			\label{fig:mot4-3}
		\end{subfigure}
	\caption{\textit{All graph-based systems fail in replacing in-DRAM
	solution. DiskANN and Starling can never meet the latency SLAs. PipeANN
	shows an insuperable gap in throughput. Srch refers to the search business, with latency SLAs of 10 ms. And Rec \& Ads refers to the recommendation and advertising, with latency SLAs of 5 ms. }}
	\label{fig:mot-graph}
\end{figure}
\begin{table}[t]
	\centering
	\resizebox{\linewidth}{!}{
		\begin{tabular}{c|ccc}
			\noalign{\hrule height 1.3pt}
			Hardware & DRAM~\cite{DDR5RDIMM} & Gen5 SSD~\cite{PM9D3A} &
			Gen4 SSD~\cite{PM9A3} \\
			\noalign{\hrule height 0.8pt}
			Price (\$/GB)     & 8  (\textbf{\textit{100\%}})        & 0.2
			(\textbf{\textit{2.5\%}})          & 0.15 (\textbf{\textit{1.9\%}})
			\\
			BW (GB/s) & $12\times38$  (\textbf{\textit{100\%}})   &
			$12\times12$  (\textbf{\textit{32\%}})  & $12\times6.5$
			(\textbf{\textit{17\%}}) \\
			\noalign{\hrule height 1.3pt}
	\end{tabular}} \caption{\textit{Price and bandwidth comparisons as of Dec. 2025.}}
	\label{tab:dram-ssd-price}
	\vspace{-3mm}
\end{table}

Recall that our CM and RAG services (see \S\ref{sec:background}) have adopted a
hybrid DRAM-SSD architecture for the graph-based DiskANN to reduce memory
consumption. Given that SSDs are becoming increasingly affordable and
high-performance as~\autoref{tab:dram-ssd-price}, it thus motivates us to consider
integrating SSDs into other services (e.g., search) as well.  Here, we explore
the feasibility by conducting comprehensive evaluations of existing graph-based
DRAM-SSD ANNS systems, including DiskANN, Starling, and PipeANN, on our standard
96-core server equipped with 12 PCIe-5.0 SSDs. Using the SIFT100M
dataset~\cite{SIFT}, we target a 90\% recall rate and evaluate top-$k$ values from 10
to 3,000, which is consistent with both the BigAnnBenchmark~\cite{BigANN} and
our online services' requirements (\textsection \ref{subsec:xhsanns}).
Unfortunately, the results are disappointing in terms of both latency and
throughput.

\noindent\textbf{Latency.}~\autoref{fig:mot4-1} and ~\autoref{fig:mot4-2} report
the mean and 99.9th percentile tail latency of the three candidates.  We test
them with a latency-friendly strategy by executing single-threaded queries
(i.e., lowest concurrency). We can observe both DiskANN and Starling
consistently fail to meet both our average and tail latency SLAs in the field
(the green shaded area). PipeANN, by leveraging intra-query parallelism (i.e.,
multi-threaded beam-search traversal) to better utilize SSD bandwidth, achieves
the lowest latency among the three but still struggles to satisfy the online
SLAs at high top-$k$ values, especially for tail latency.

\begin{figure}[t]
	\centering
	\includegraphics[width=0.99\linewidth]{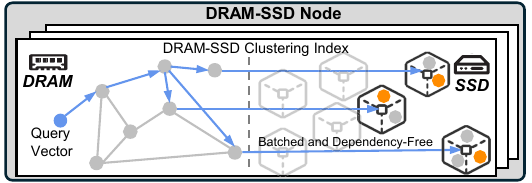}
	\centering
	\caption{\textit{DRAM-SSD clustering-based index follows the batched and
	dependency-free I/O pattern on the NVMe SSDs.}}
	\label{fig:clustering}
		\vspace{-3mm}
\end{figure}

\noindent\textbf{Throughput.} To further explore whether PipeANN can serve as a
(partial) replacement, we evaluate its throughput against in-DRAM HNSW under the
latency SLAs.  As shown in \autoref{fig:mot4-3}, even with 12 PCIe-Gen5 SSDs
providing over 30\% of DDR5 DRAM bandwidth (\autoref{tab:dram-ssd-price}),
PipeANN's peak throughput remains 10–25$\times$ lower than that of HNSW. The
root cause is that the graph-based DRAM-SSD ANNS inherently suffers from
strongly serialized, dependency-chain I/Os during search (i.e., each expansion
step depends on the results of previous reads).  This serialized access pattern
magnifies the raw latency gap between SSDs and DRAM (around $10^2$-$10^3$). In
our low-latency and high-throughput online services, this limited performance
rules out the possibility of graph-based DRAM-SSD systems as a viable
replacement for in-DRAM solutions at scale.

\subsection{Clustering-Based ANNS Brings Hope}
Failing to port in-memory ANNS with graph-based hybrid designs motivates us to
reconsider another design, clustering-based DRAM-SSD ANNS. At a high level,
\autoref{fig:clustering} shows that SPANN~\cite{SPANN}, a representative
clustering-based ANNS, uses a clustered layout. The clusters' centroids are kept
in DRAM, while vectors are stored on SSD as clustered posting lists. During
search, it first searches the in-memory graph of centroids to identify the
$nprobe$ number of (e.g., $3$ in \autoref{fig:clustering}) closest partitions, then issues
batched reads to fetch the corresponding posting lists, and finally computes the
top-$k$ neighbors from the loaded candidates of these lists. 

\begin{figure}[t]
	\centering
	\begin{subfigure}[t]{0.48\linewidth}
		\includegraphics[width=\linewidth]{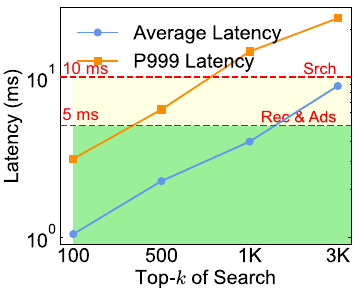}
		\caption{\textit{Near-SLAs latency.}}
		\label{fig:mot-sp1}
			\vspace{-3mm}
	\end{subfigure}
	\centering
	\begin{subfigure}[t]{0.48\linewidth}
		\includegraphics[width=\linewidth]{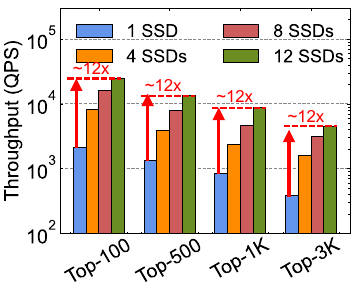}
		\caption{\textit{Scalable throughput.}}
		\label{fig:mot-sp2}
			\vspace{-3mm}
	\end{subfigure}
	\centering
	\caption{\textit{Based on high-bandwidth modern SSDs, the clustering-based
			SPANN shows nearly qualified search latency and scalable throughput
			under the large top-k values.}}
	\label{fig:mot-sp}
	\vspace{-3mm}
\end{figure}
Historically, this line of research was known to be latency-friendly but largely
dismissed in the field due to the limited
throughput~\cite{FusionANNS,Starling,PipeANN}. However, the recent advancement
in high-bandwidth NVMe SSDs (e.g., 12\,GB/s from Gen5 SSD) has changed this
landscape. As shown in ~\autoref{fig:mot-sp}, using SPANN and up to 12 PCIe-Gen5
SSDs on SIFT100M, our evaluations demonstrate that, with the batched and
dependency-free I/O pattern, this clustering-based design can already approach
latency SLAs of online services under processing single-threaded query and
achieve scalable throughput by adding SSDs. In addition, \autoref{fig:mot-sp1}
shows that, even when we increase top-$k$ up to $3\times10^3$, SPANN nearly
keeps the average and tail latency within or close to the SLAs of search,
recommendation, and advertising.  Moreover, \autoref{fig:mot-sp2} illustrates
that the throughput scales almost linearly with the number of SSDs and reaches
$\sim12\times$ speedup when using 12 SSDs, reinforcing our assumption that
modern SSDs with clustering-based index can be a promising direction.

\begin{figure}[t]
	\centering
	\begin{subfigure}[t]{0.49\linewidth}
		\includegraphics[width=\linewidth]{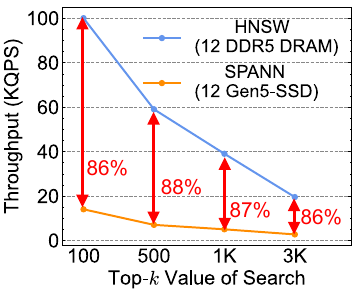}
		\caption{\textit{{Insufficient throughput.}}}
		\label{fig:cha1}
	\end{subfigure}
	\centering
	\begin{subfigure}[t]{0.49\linewidth}
		\includegraphics[width=\linewidth]{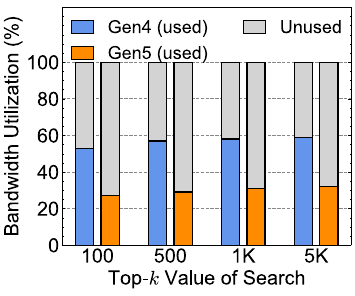}
		\caption{\textit{{Low bandwidth utilization.}}}
		\label{fig:cha2}
	\end{subfigure}
	\begin{subfigure}[t]{0.49\linewidth}
		\includegraphics[width=\linewidth]{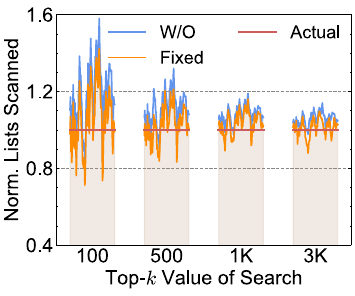}
		\caption{\textit{{Un-adapted range of scans.}}}
		\label{fig:cha3}
	\end{subfigure}
	\begin{subfigure}[t]{0.49\linewidth}
		\includegraphics[width=\linewidth]{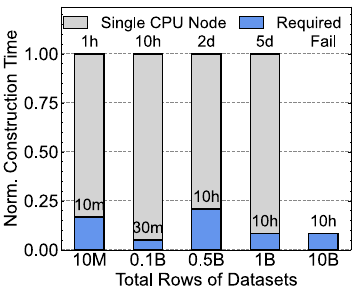}
		\caption{\textit{{Long-time index building.}}}
		\label{fig:cha4}
	\end{subfigure}
	\centering
	\caption{\textit{Clustering-based SPANN still has deficiencies. Its
	performance also fails to fully meet the throughput SLAs ($\sim$12\%-14\%
	of in-DRAM HNSW), due to the low bandwidth utilization and un-adapted range
	of scans. Furthermore, its single-node CPU-only construction of indexes 
	incurs a heavy expenditure of time, violating the freshness of services.}}
	\label{fig:cha}
	\vspace{-3mm}
\end{figure}
\subsection{Challenges in Deploying SPANN at Scale}
However, even when we pair SPANN with the array of NVMe SSDs, such a
solution still cannot be directly deployed in the field. Challenges persist in
both online search and offline construction. For online serving, the achieved
search throughput is well below our production SLAs. Moreover, in the offline
stage, the construction of indexes incurs prohibitive time overheads that must
be substantially reduced.

First, SPANN cannot yet serve as a scalable replacement for the in-DRAM HNSW
indexes deployed in our online stack. Under the latency SLAs of online services,
even on a 96-core node with 12 PCIe 5.0 SSDs, the throughput remains
insufficient. As shown in \autoref{fig:cha1}, there is still a throughput gap of about 86-88\%
between SPANN and the in-DRAM HNSW baseline.\footnote{Referring to our
production index-building settings, \texttt{edges} and \texttt{efConstruction}
of HNSW are set as 24 and 120. SPANN is set as \S\ref{sec:setup}} This
gap is largely due to insufficient utilization of NVMe SSDs' bandwidth.
\autoref{fig:cha2} shows that SPANN uses only $\sim$26-59\% of the available
bandwidth on PCIe Gen5 and Gen4 arrays, highlighting a primary opportunity to
scale throughput by driving SSD bandwidth actually utilized closer to the
hardware limits.

Second, SPANN’s current strategy for determining the scan range (i.e., the
number of loaded clusters $nprobe$) impacts both throughput and search quality.
It relies on a fixed distance-based pruning rule shown in
~\autoref{eq:fixed_prune}. After locating the closest centroid $c_{i1}$, SPANN
includes clusters $\mathbf{X}_{ij}$ in the search if distances from $\mathbf{q}$
to their centroids $c_{ij}$ are within a $(1+\epsilon)$ factor of the distance
to $c_{i1}$, where $\epsilon$ is chosen empirically and does not adapt to
query difficulty or data distribution. \autoref{fig:cha3} presents statistics
for $\sim$100 queries under varying top-$k$. The scanned range after
pruning is only slightly smaller than the no-pruning (W/O) baseline, yet for
each individual query, it often overshoots the range needed to reach the target
recall (e.g., 90\%), wasting I/O and reducing throughput; or undershoots it,
causing large variance in recall and resulting in unstable search quality. 
\begin{equation}\small
	\label{eq:fixed_prune}
	\begin{aligned}
		\mathbf{q} \xrightarrow{\text{search}} \mathbf{X}_{ij}
		&\Longleftrightarrow
		\operatorname{Dist}(\mathbf{q}, c_{ij})
		\le (1+\epsilon)\, \operatorname{Dist}(\mathbf{q}, c_{i1}), \\
		&
		\operatorname{Dist}(\mathbf{q}, c_{i1}) \le \operatorname{Dist}(\mathbf{q}, c_{i2})
		\le \cdots \le \operatorname{Dist}(\mathbf{q}, c_{iK})
	\end{aligned}
\end{equation}

\begin{figure*}[t]
	\centering
	\includegraphics[width=\linewidth]{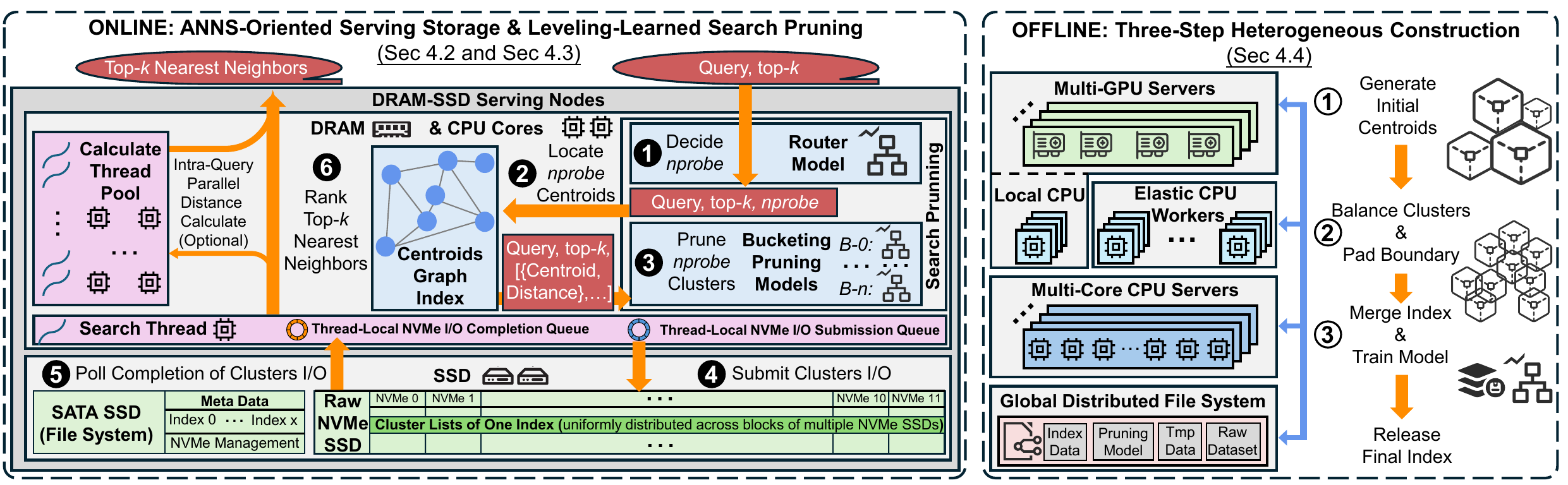}
	\centering
	\caption{\textit{Design overview of {\sysname} with workflows of online
	search and offline index construction.}}
		\label{fig:overview}
		\vspace{-3mm}
\end{figure*}

Finally, SPANN also faces significant challenges in offline index
(re)building. As shown in \autoref{fig:cha4}, the existing implementation
constructs the clustering structures on a standalone CPU node.  When the dataset
grows from ten million to tens of billions of vectors, the construction time
escalates from several hours to multiple days. These construction windows are
far longer than what is acceptable in our production scenarios, where
embedding models and vector data are refreshed frequently. Consequently, index
building also becomes a major bottleneck hindering the practical deployment of
SPANN.

\section{\sysname Design}

\subsection{Overview}
We present \sysname, a high-performance and cost-effective ANNS system using
clustering-based indexes on all-flash servers.  Figure~\ref{fig:overview}
presents an overview of \sysname, including online serving (i.e., the left
half of Figure~\ref{fig:overview}) and offline index construction (i.e., the
right half). 

\noindent\textbf{Online serving.} We employ all-flash servers as ANNS serving
nodes where each is equipped with 12$\times$2\,TB PCIe Gen5 NVMe SSDs,
12$\times$96\,GB DDR5 DRAM, and a 96-core CPU. Recall that existing solutions
such as SPANN fall short in throughput due to bandwidth utilization and pruning
efficiency. We therefore redesign the storage stack and the pruning module. The
key idea is eliminating overhead from I/O software and introducing adaptive
learning-based models. At a high level, for each index released to serving, we
use DRAM to store the centroid graph for locating nearest clusters and weights
of the search pruning module (router and pruning). Meanwhile, the cluster lists
are striped across raw NVMe SSDs as the granularity of logical blocks, bypassing
the traditional Linux software stack.  Since a single node can usually host
multiple vector indexes, we use an extra SSD to store the metadata of all
indexes and management of raw SSDs. 

For each query with target top-$k$, \circleNum[0.8em]{1} the router model decides the
$nprobe$ value, \circleNum[0.8em]{2} then \sysname uses the centroids graph index to locate the
$nprobe$ closest centroids, and \circleNum[0.8em]{3} the leveling pruning model routed
at the first step further prunes the candidate clusters. \circleNum[0.8em]{4} The
search threads submit asynchronous NVMe I/O commands for the selected cluster
lists, \circleNum[0.8em]{5} poll completions of cluster reads on the hardware completion
queue, and \circleNum[0.8em]{6} calculate the loaded vectors in the local thread or
dispatch them to the calculate thread pool to compute distances and finally rank
the top-$k$ nearest neighbors as the result of ANNS.

\noindent\textbf{Offline index construction.} To facilitate frequent index
building, {\sysname} employs multi-GPU servers to form a three-stage pipeline.
Note that a major drawback in previous builds by a single node's CPUs is limited
computing power and the inability to scale.  Hence, we introduce GPUs to
generate initial centroids and distributed nodes to accelerate balancing and
padding. Backed by a global distributed file system holding the final index,
pruning models, temporary data (e.g., checkpoints), and raw datasets, we
orchestrate the entire construction pipeline end-to-end, achieving minute-level
index building for datasets up to million-scale vectors and hour-level construction for
billion-scale datasets.

In offline construction, \circleOne[0.8em]{1} multi-GPU servers run k-means to generate
initial centroids from the raw dataset and save these centroids to the
distributed storage. \circleOne[0.8em]{2} Then, owning initial centroids, local CPUs of
multi-GPU servers or elastic CPU workers split and rebalance clusters from the
initial centroids, pad cluster boundaries, and write intermediate index shards
to the global distributed file system. \circleOne[0.8em]{3} Multi-core CPU servers merge
the shards, build the graph for all final centroids, train leveling pruning
models, and finally materialize the index files for releasing them on serving
nodes.

\noindent\textbf{Design advantages.} With these designs, \sysname effectively
solves the aforementioned challenges. First, as \circleNum[0.8em]{4} and \circleNum[0.8em]{5} of
the online serving in the left of~\autoref{fig:overview}, it bypasses the kernel
on the throughput-critical I/O path (i.e., cluster-list reads) completely to
exploit the high bandwidth of modern multi-SSD arrays, minimizing software
overhead by directly steering NVMe SSDs to match ANNS access patterns
(\textsection \ref{sec:storgae}). 

Second, as \circleNum[0.8em]{1}, \circleNum[0.8em]{2}, and \circleNum[0.8em]{3} of the online serving,
it adaptively chooses the search range (i.e., probed clusters) on a per-query
basis, achieving target recall while avoiding unnecessary I/O and computation: a
trained router model predicts the recall-sufficient $nprobe$, and leveling
pruning models further eliminate unnecessary scans (\textsection
\ref{sec:prune}). 

Third, as \circleOne[0.8em]{1} and \circleOne[0.8em]{2} of the offline building, to support
frequent index building for incremental embedding model training and
continuously evolving vector data, we utilize heterogeneous acceleration and
make support for elastic scaling.  Multi-GPU servers accelerate centroid
generation on large-scale datasets and we can dynamically allocate CPUs to
speed up the following fine-grained index construction flexibly according to
requirements (\textsection \ref{sec:build}).

\begin{figure}[t]
	\centering
	\begin{subfigure}[t]{0.48\linewidth}
		\includegraphics[width=\linewidth]{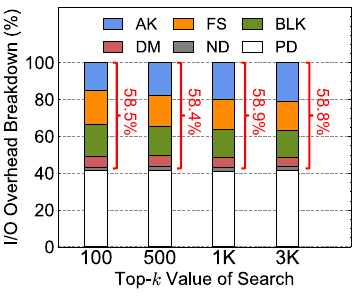}
		\caption{\textit{I/O breakdown of SPANN.}}
		\label{fig:d-11}
	\end{subfigure}
	\centering
	\begin{subfigure}[t]{0.48\linewidth}
		\includegraphics[width=\linewidth]{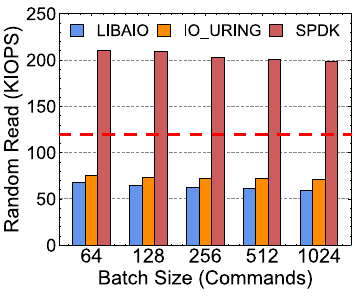}
		\caption{\textit{Ideal IOPS  per core.}}
		\label{fig:d-12}
	\end{subfigure}
	\centering
	\caption{\textit{SPANN’s I/O path relies on the traditional Linux I/O
			software stack, including \textit{AK} (application-kernel
			switching), \textit{FS} (file system), \textit{BLK} (block layer),
			\textit{DM} (device mapper for RAID), and \textit{ND} (NVMe driver).
			By issuing fixed-size 12 KB reads in batches—a new batch only after
			the previous one completes—we show the I/O breakdown of SPANN, and compare the performance of \texttt{libaio},
			\texttt{io\_uring}, and SPDK.}}
	\label{fig:d-1}
	\vspace{-3mm}
\end{figure}
\subsection{ANNS-oriented Storage Stack}
\label{sec:storgae}
To better serve the clustering-based ANNS indexes, we study the existing storage
stack and propose an ANNS-oriented storage design. We start by analyzing the I/O
behavior of clustering-based search and then discuss the corresponding design
choices which take advantage of multiple NVMe SSDs with the help of the user-space
SPDK driver~\cite{SPDK}.

\noindent\textbf{Understanding I/O patterns.} Clustering-based ANNS has two
outstanding I/O patterns during online search. First, the search process tends
to generate a large number of batched reads. For example, a single query can
generate up to $\sim10^3$ cluster-list loading requests when searching the
top-$3{,}000$ nearest neighbors, a typical case in online search and
advertising. Such massive reads can lead to a demandingly high IOPS requirement.
Taking SIFT100M as an example, under a 10\,ms latency SLA, computation (e.g.,
finding the nearest centroids, distance calculations on vectors loaded, and
ranking top-$k$ results) can already cost $\sim$2-4\,ms with only $\sim$6-8\,ms
left for I/Os, which in return translates to $\sim$120-170\,KIOPS per search
thread on a single core. However, the I/O stack used in SPANN can often only
achieve $\sim$30-40 KIOPS. 

Second, the clusters of an index have the same size which is also the size of the
issued reads. This is because, to avoid long-tail latency and the boundary error
during search~\cite{SPANN}, the clustering-based index usually balances
clusters' sizes below a threshold and pads clusters with boundary vectors to the
same size (e.g., 12\,KB per cluster for SIFT in SPANN).  As a result, each
cluster can have same count of pages, and each probe typically issues
fixed-size reads (e.g., 12\,KB from three 4KB pages) when loading the nearest clusters.

\noindent\textbf{Exploring design choices.} We next profile the I/O overhead of
SPANN (~\autoref{fig:d-11}) and three popular I/O stacks including
\texttt{libaio}, \texttt{io\_uring}, and SPDK (~\autoref{fig:d-12}) under
workloads of batched and fixed-size read patterns (i.e., the same I/O behavior
as the clustering-based search) as shown in~\autoref{fig:d-1}.  Results indicate
that the overhead of the traditional I/O stack with \texttt{libaio} dominates the
end-to-end path (up to 58\% of the total) across all top-$k$ settings, even
exceeding that of accessing the SSD (i.e., \textit{PD}, physical devices). 
\texttt{io\_uring} shows moderate improvements over \texttt{libaio} since it
does not have system calls~\cite{ModernNVMe}.  SPDK achieves the highest IOPS by
bypassing the traditional kernel stack completely.  Note that these throughput numbers
are all measured without considering the computation cost in~\autoref{fig:d-12}.
Given that \texttt{io\_uring} and \texttt{libaio} have already failed to meet
the expected IOPS, we therefore choose to build our serving storage over SPDK. 
\begin{figure}[t]
	\centering
	\includegraphics[width=0.99\linewidth]{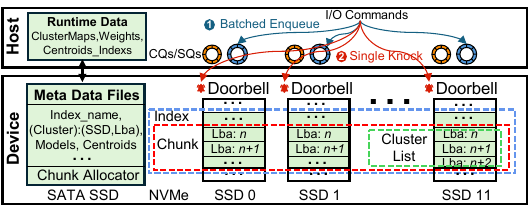}
	\centering
	\caption{\textit{Meta data as files, clusters' lists as raw blocks.}}
	\label{fig:storage-detail}
	\vspace{-5mm}
\end{figure}

\noindent\textbf{Customizing the storage stack.} Adopting SPDK means  dropping the
support from the kernel, and we hence need to manage the data layout of indexes, the
I/O control of raw devices, and the multi-device space allocation.  We
demonstrate our serving storage stack, as shown in ~\autoref{fig:storage-detail}.

\begin{itemize}[leftmargin=*, nosep]
	\item \textit{Data layout.} The index data consists of metadata and cluster
	lists.  Metadata includes the index name, mappings of each cluster and its
	physical location (SSD id and LBA), pruning models, and the centroid index.
	Since these structures are small and can reside in memory without external
	reads at runtime, we simply store them as regular files on a local SSD. In
	contrast, the efficiency-critical cluster lists are placed directly on the
	logical blocks of the NVMe SSD array. Each cluster list occupies a
	contiguous block range of a single SSD (e.g., a cluster list uses three
	consecutive LBAs of SSD 11 in~\autoref{fig:storage-detail}), so that reading
	one cluster only requires a single I/O command, avoiding multiple I/Os due
	to crossing devices or blocks.
	\item \textit{I/O control.} For the only runtime external I/O--loading
	cluster lists--we bypass system calls and directly submit NVMe commands to
	hardware queues while polling for completions. With this direct control of
	devices, we optimize submission at hardware granularity. I/O commands are
	enqueued to host-side NVMe queues in batches, and each NVMe device is
	notified with a single PCIe doorbell knock per batch. This eliminates
	hundreds of per-command PCIe round-trips (at the microsecond scale) and
	significantly reduces CPU overhead on the critical read path.
	\item \textit{Space allocation.} Since all clusters are padded to a fixed
	size, we can pre-allocate cluster-aligned regions on SSDs, avoiding
	fragmentation and complex file-system allocators. Hence, on raw NVMe
	devices, we exploit this property with a unified chunk-based free-list
	allocator (e.g., 64\,MB per chunk) that manages SSD space for all indexes.
	It allocates and recycles fixed-size regions for deploying and deleting indexes at
	the chunk granularity. Each index then partitions its chunks into
	consecutive block ranges sized to a cluster list and assigns each range to a
	single cluster.
\end{itemize}

\subsection{Leveling-learned Search Pruning}
\label{sec:prune}
\textbf{Problem and purpose.}
In clustering-based search, the number of probed clusters (i.e., $nprobe$)
determines recall and performance.  The trade-off is that an overly large
$nprobe$ significantly increases the I/O and computation cost due to redundant
scans, while an overly small $nprobe$ hurts recall.

\noindent\textbf{Adapting nprobe by pruning.} Many prior works such as
Quake~\cite{Quake}, Auncel~\cite{Auncel}, and LAET~\cite{LAET} propose early
termination in pruning, which seems suitable for our adaptive adjustment of
$nprobe$. However, they in fact cannot be applied to our scenario due to the lack of
support for changing top-$k$ of requests and the strong dependence on
intermediate results. First, existing methods typically make early-termination
rules for a fixed top-$k$ while varying target recall. In contrast, in our
services, the recall target can be predetermined from scenarios (e.g., 90\% in
online search) but top-$k$ can vary across requests (e.g., 100-3,000 in
online search).  

Second, LAET also relies on intermediate features such as the distances to the
current 1st and 10th neighbors after probing a subset of the nearest candidate
clusters located. Similarly, Quake and Auncel need to check after each newly
scanned cluster whether the recall target is met. Such iterative
probe–compute–decide loops serialize cluster I/Os, and thus are unfriendly to the
batched SSD reads for cluster lists, leading to reduced bandwidth utilization.

\noindent\textbf{Leveling-learned search pruning.} We propose the
leveling-learned search pruning (LLSP) based on gradient boosting decision
trees (GBDT)~\footnote{GBDT is an effective architecture for pruning of online
search because of its fast training (e.g., minute-level overhead for 1 million
entries), low-overhead inference (e.g., $\sim$10-30 us per prediction), and
small memory footprint (e.g., only hundreds of KB per
model).}~\cite{GBDT,LightGBM}.  First, to address varying top-$k$ values, LLSP
has a router model for choosing the maximum range according to the different
search difficulty (i.e, the distribution of query vector and the top-$k$ value)
of requests. Second, it has a group of multi-level learning-based models for
pruning redundant scans. At this stage, to keep batched clusters reads, we avoid
using intermediate search results as features, and instead use only pre-search
information: the query vector, the top-$k$ value, and statistics of its
distances to all nearest candidate centroids~\footnote{Feature-importance
analysis shows query vectors and centroid-distance ratios can contribute nearly
50–70\% of predictive power~\cite{LAET}.}.

\begin{figure}[t]
	\centering
	\includegraphics[width=0.98\linewidth]{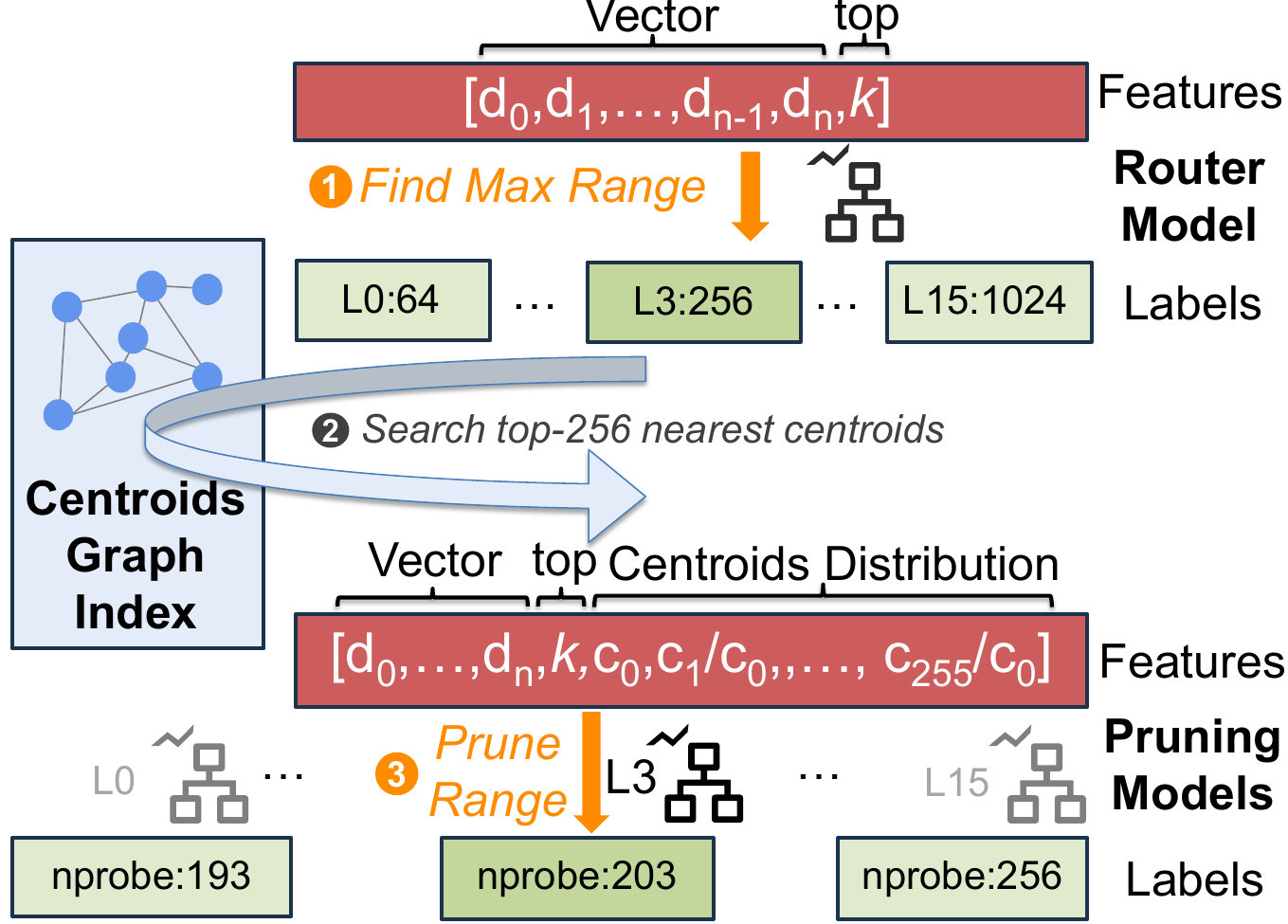}
	\centering
	\caption{\textit{Pruning workflow of search during online serving.}}
	\label{fig:prune-detail}
	\vspace{-3mm}
\end{figure}
\noindent\textbf{Online-offline workflow.} In detail, we first present the
pruning process of search in online serving and then describe how the models are
trained in offline building.

\begin{itemize}[leftmargin=*, nosep]
\item \textit{Online serving workflow.} As shown in~\autoref{fig:prune-detail},
we first feed the query vector and its requested top-$k$ into the router model,
which predicts a coarse maximum search range's level (e.g., level 3 with 256 as
$nprobe$ here, L3:256) as the upper bound of $nprobe$. Then the centroids graph
index returns the top-$nprobe$ nearest centroids. Finally, LLSP constructs
pruning features from the query, top-$k$, and the centroid–distance distribution
(the nearest centroid-query distance and relative ratios of the following 255
centroids' to the 1st centroid's), and applies the corresponding level-specific
(i.e., level 3 here) pruning model to refine $nprobe$ for batched loading
nonredundant clusters  from SSDs.

\item \textit{Offline training workflow.} We first set a series of range levels
with increasing upper bounds on $nprobe$ (e.g., 64 to 1,024 with a step of 64).
From a recent time window (e.g., the previous 1 day for search, 1 hour for
recommendation and ads), we uniformly sample about 1\% of logged items as the
training supervision. This is because most production traces show up to 90\%
duplication in short windows~\cite{ReCANet,RepeatExplain,RepeatTemporal}. To
avoid the overhead of brute-force ground truth, we approximate labels by running
non-pruning search with a large $nprobe$ (e.g., 4{,}096). We then train the
router model by the procedure that, for each query and business-required
top-$k$, finding the smallest level whose range meets the target recall and
using the pair (query, top-$k$) as features and the level as label. Afterward,
within each level, we derive labels for training pruning models by gradually
decreasing $nprobe$ from the level’s maximum until recall drops to the
threshold; the resulting $nprobe$ is taken as the label. And the query, top-$k$,
and centroid-distance distributions under the maximum $nprobe$ form the feature
vectors together. With the labels (i.e., actual $nprobe$ minimized) and these
features, we train the pruning model of each level.

\end{itemize}

\subsection{GPU-accelerated and Elastic Building}
\label{sec:build}
\begin{figure}[t]
	\centering
	\includegraphics[width=0.99\linewidth]{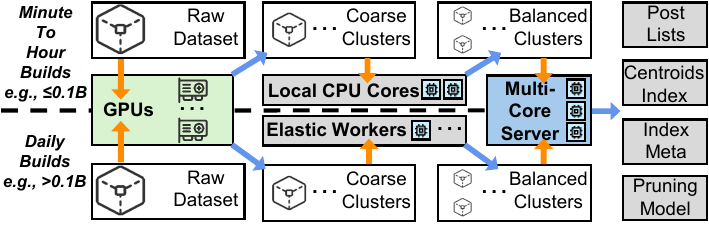}
	\centering
		\vspace{-1mm}
	\caption{\textit{Pipelines of the three-step index construction.}}
	\label{fig:build-detail}
	\vspace{-3mm}
\end{figure}

To reduce the index construction time for million- to billion-scale datasets from days to tens of minutes, we leverage  GPU-based acceleration and further design an elastic
pipeline to scale the computation from a distributed CPU pool.

\noindent\textbf{Procedure and issues of existing index construction.} Again, we
first examine the status quo (i.e., the index construction in SPANN).  There are
three steps.  First, it applies hierarchical k-means to partition vectors into
many size-bounded clusters, avoiding oversize and unbalanced partitions. Then,
it performs closure multi-cluster assignment that duplicates boundary vectors,
using RNG rules~\cite{RNGRule1980} to control redundant copies.  Finally, it
builds an in-DRAM graph on the centroids of all clusters for locating the
nearest clusters to the query during online serving.

We can see that there are two main drawbacks in the first clustering step, which
incurs the main overhead (up to $\sim$60-80\%) of existing constructions.
First, k-means on large-scale datasets incurs repeated distance calculations on
high-dimensional vectors and thus places a heavy load which is beyond the
capability of a CPU. Second, when vectors increase to billions from millions, it
still requires multiple hours of building, requiring paralleling about
$10^2$-$10^3\times$ resources as data scaling. However, the current SPANN
construction can only use one CPU in building and thus takes days to finish building a
billion-scale index, which is unacceptable for production services.  

\noindent\textbf{GPU acceleration and elastic scaling.} Hence, we exploit GPUs
to accelerate centroid initialization and iterative balancing, and employ the
elastic CPU pool to address the scaling of datasets.
~\autoref{fig:build-detail} shows the workflow of our three-step index
construction pipeline.

\begin{figure}[t]
	\centering
	\includegraphics[width=0.99\linewidth]{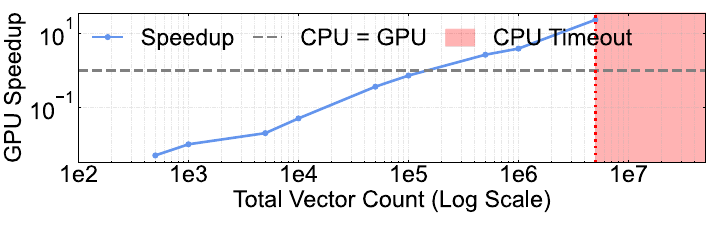}
	\centering
	\vspace{-2mm}
	\caption{\textit{The speedup ratio of GPU clustering across varying
			vector counts with an NVIDIA L20 vs. 48 CPU cores.}}
	\label{fig:clustering-ratio}
	\vspace{-3mm}
\end{figure}

 \textit{Generating coarse centroids with GPUs.} First, for datasets of all
	scales, we use GPUs to perform coarse clustering. In this step, we do not
	directly split clusters down to their final size, because we found that GPU
	acceleration is not always effective{\footnote{At present, the development our GPU K-means is based on Faiss.}}. As shown
	in~\autoref{fig:clustering-ratio}, GPUs provide up to orders-of-magnitude speedups on multi-million–scale and more vectors, while
	for small jobs (e.g., clustering on <$10^5$ 128-dim vectors), extra
	host–device transfers can dominate, making GPUs slower than the CPUs.

\textit{Constructing balanced posting lists.}
After the early clustering phase on GPUs, most clusters are already smaller than
the threshold (e.g., $10^5$ by default). We then use CPUs to perform
fine-grained splitting and redundant padding of boundary vectors. Depending on
the application scenario and dataset scale, we adopt two execution schemes. 

For minute-to-hour builds of common-scale datasets (e.g., $<0.1$B vectors from
recommendation and advertising), we employ local CPU cores to immediately finish
the fine splitting and duplication with RNG checking, avoiding network transfers
and extra scheduling overhead. 

For search services that require daily rebuilds on larger corpora, elastic
workers in online CPU clusters perform the fine-grained clustering and padding during
off-peak hours, opportunistically harvesting idle resources from existing
clusters under diurnal load patterns (i.e., lighter utilization at night and
heavier loads during the day).

In addition, to avoid impacting real-time services, we enforce a QoS policy:
whenever online and offline jobs contend for resources, online traffic always
has priority and the index-building task on that node is preempted and
terminated, and would retry later. To control the resulting tail latency, we
further introduce task re-assignment and node eviction: once a task exceeds a
retry threshold, it is reassigned to another node, and the original node is
temporarily removed from the resource pool, preventing a few unstable nodes from
dominating the overall construction time.

\textit{Building final index data.} Finally, balanced and replicated clusters
generated by both local CPUs of GPU servers and elastic CPU workers are
uniformly prepared for releasing.  They are consolidated on the multi-core
servers to produce deployable index files, including posting lists, centroid index,
metadata, and the pruning model.
	
\section{Evaluation}
In this section, we conduct extensive experiments to evaluate the practical impacts of {\sysname}, summarized as follows:
\begin{itemize}[leftmargin=*, nosep]
	\item {\sysname} achieves 2-16$\times$ throughput of baselines, and can meet latency SLAs to replace in-DRAM HNSW (\textsection \ref{sec:performance-eva}).
	\item {\sysname} achieves 1.6-7.5$\times$ SSD bandwidth utilization of other DRAM-SSD systems, yielding up to 87\% improvement by simply upgrading SSDs (\textsection \ref{sec:storage-eva}).
	\item 
	Under the same average recall, \sysname ensures that >80\% of queries individually meet the target recall by near 30 percentage points over SPANN (\textsection \ref{sec:accuracy-eva}).
	\item GPU acceleration enables constructing 0.1B-vector indexes within 1 hour. And by elastically scaling resources, 10B vectors can be built within 10 hours (\textsection \ref{sec:build-eva}).
	\item {\sysname} can reduce DRAM usage by over 90\% and improve throughput-per-dollar by 8.3$\times$ for the serving deployment (\textsection\ref{sec:cost-eva}).
	
\end{itemize}

\subsection{Setup}
\label{sec:setup}
\begin{table}[h]
	\centering
	\small
	\resizebox{0.95\linewidth}{!}{
		\begin{tabular}{lcccc}
			\toprule
			\textbf{Dataset} & Scale & Dim. & Total size & Top-$k$ Range \\
			\midrule
			SIFT              & 0.1B\&10B  & 128  & 12\,GB\&1.2\,TB & 10–3{,}000 \\
			Red\textsf{Srch} & 0.5B\&10B  & 64   & 30\,GB\&0.6\,TB & 100–3{,}000 \\
			Red\textsf{Rec}  & 0.1B         & 64   & 6\,GB            & 100–1{,}000 \\
			Red\textsf{Ads}  & 20M          & 128  & 2.5\,GB           & 100–3{,}000 \\
			Red\textsf{CM}   & 0.1B         & 64   & 6\,GB             & 100–500 \\
			Red\textsf{RAG}  & 4M           & 1024 & 4\,GB             & 10–100 \\
			\bottomrule
		\end{tabular}
	}
	\vspace{-1mm}
	\caption{Statistics of the evaluated datasets.}
	\label{tab:datasets}
	\vspace{-3mm}
\end{table}
\noindent\textbf{Workloads.} We evaluate on the public SIFT and five production datasets, as shown in ~\autoref{tab:datasets}. SIFT10B is created by 10$\times$ replicating SIFT1B. Real datasets span 4M–10B vectors from our main services mentioned. Similarity metric is the L2 distance. For SIFT, we uniformly generate query top-$k$ values ranging from 10 to 3,000, while for other datasets, top-$k$ values are sampled from our real production traces. 
\begin{figure}[t]
	\centering
	\begin{subfigure}[t]{\linewidth}
		\includegraphics[width=\linewidth]{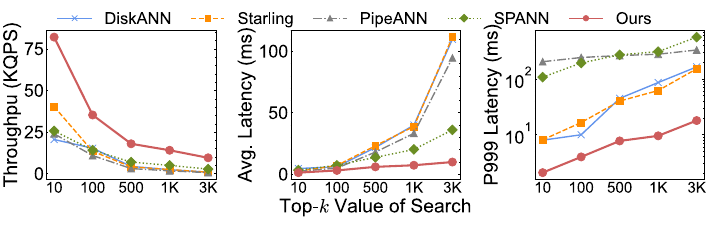}
		\vspace{-5mm}
		\caption{\textit{Performance across various top-$k$ of search, under recall = 90\%.}}
		\label{fig:eva1-1}
	\end{subfigure}
	\centering
	\begin{subfigure}[t]{\linewidth}
		\includegraphics[width=\linewidth]{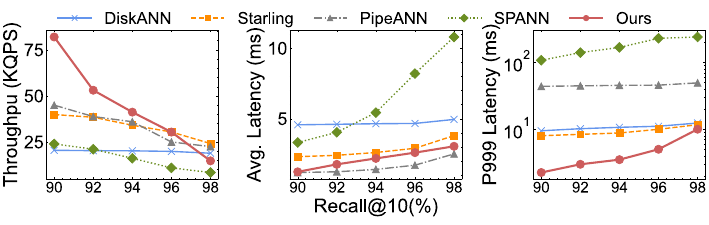}
				\vspace{-5mm}
		\caption{\textit{Performance as the top-10 search, under various recall rates.}}
		\label{fig:eva1-2}
	\end{subfigure}
	\centering
	\caption{\textit{Comparisons about throughput, average and tail latency under different top-k and recall, using SIFT100M.}}
	\label{fig:eva-1}
			\vspace{-3mm}
\end{figure}

\noindent\textbf{Baselines.} 
We compare {\sysname} with DiskANN, Starling, PipeANN, SPANN, and HNSW. For DRAM-SSD baselines, we use their released implementations. We tune beam widths of graph-based systems (16 for DiskANN and Starling, 32 for PipeANN, unless otherwise specified) to achieve their optimal performance, since multi-SSD setups offer sufficient bandwidth. We set DRAM budget as 25\% of the size of raw datasets~\cite{PipeANN}. For \sysname and SPANN,  we reduce the replication factor to 4, with centroids accounting for 8\% of the total scale, thereby setting the DRAM:SSD ratio to 1:20 to match the hardware capacity. We include in-DRAM HNSW that is widely used in production for low-latency services, which serves as a strong performance reference for purely memory-resident deployments but with higher DRAM cost.

\noindent\textbf{Specifications.} All DRAM–SSD tests are conducted on  our testbed equipped with the 96-core AMD EPYC 9654 CPU, 12$\times$96 GB DDR5 DRAM, 12$\times$1.92 TB PCIe-5.0 NVMe SSDs. In-DRAM distributed HNSW uses our standard configuration with 32-core CPUs and 256 GB DDR5 DRAM to hold shards, a configuration commonly adopted in production.
 \begin{figure}[t]
	\centering
	\includegraphics[width=\linewidth]{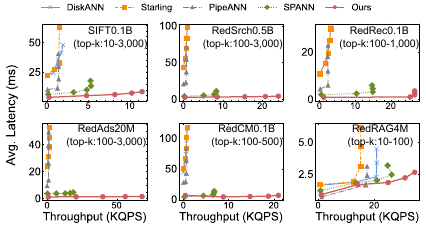}
	\caption{\textit{Trends of throughput and average latency.}}
	\label{fig:eva1-3}
	\centering
	\vspace{-5mm}
\end{figure}

\subsection{Search Performance}
\label{sec:performance-eva}
\noindent\textbf{End-to-end performance.} Overall, \sysname achieves the best end-to-end efficiency across a wide range of top-$k$ and recall targets, as shown in~\autoref{fig:eva-1}.

\textit{Varying top-$k$} (\autoref{fig:eva1-1}). With recall fixed at 90\%, \sysname sustains the highest throughput on SIFT0.1B across all top-$k$ values. As top-$k$ increases from 10 to 3{,}000, its throughput degrades more slowly than that of other DRAM-SSD systems, while average latency stays within 10 ms and P99.9 latency remains well below 20 ms. For graph-based baselines (DiskANN, Starling, PipeANN), larger top-$k$ forces longer greedy walks due to larger candidate lists. Starling improves top-10 performance via optimized layout but is less effective for large top-$k$. The reason is that, as top-k increases, the search depth of graph-based search grows significantly. For example, for top-1000 on the SIFT0.1B, both the candidate queue length and the number of search hops increase substantially to around 1,500 or more. The number of accessed SSD pages also rises significantly, which leaves little room for gains from layout reordering of Starling.
Meanwhile, SPANN is constrained by the traditional I/O software stack. Consequently, both throughput and latency for baselines deteriorate rapidly.

\textit{Varying recall} (\autoref{fig:eva1-2}). Fixing top-$k$ at 10, \sysname delivers competitive throughput at 90-96\% recall and keeps clear advantages when recall enters the high-accuracy regime. Both average and tail latency grow smoothly with the recall, and \sysname maintains sub-10 ms average latency and the lowest P99.9 latency across the entire recall range.

\noindent\textbf{Throughput vs. latency.} We evaluate workloads from generated SIFT and productions with service-specific top-$k$ and a 90\% recall target, ramping threads until all systems saturate.

\textit{Throughput vs. average latency.} As shown in \autoref{fig:eva1-3}, graph-based systems can incur $\sim$120 ms average latency at only 0.5-3 KQPS because large top-$k$ requires long greedy walks for longer length of candidate lists (e.g., 4{,}000 for top-3{,}000). SPANN benefits from high-bandwidth SSDs, and \sysname further pushes the frontier, achieving up to 30$\times$ throughput while keeping average latency within 5-10 ms.

\begin{figure}[t]
	\includegraphics[width=\linewidth]{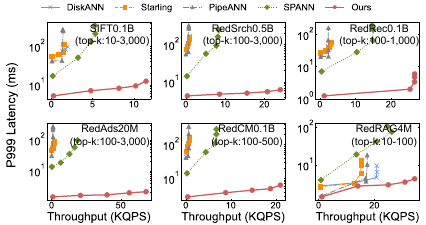}
	\caption{\textit{Trends of throughput and P999 tail latency.}}
	\label{fig:eva1-4}
\end{figure}
\textit{Throughput vs. tail latency.} For P99.9 latency (\autoref{fig:eva1-4}), graph-based systems and SPANN exhibit steep tail blow-ups once saturated. By clustering-based search and our customized I/O stack, \sysname maintains up to an order-of-magnitude lower P99.9 latency, keeping it around $\sim$10 ms.

\begin{figure}[t]
	\centering
	\includegraphics[width=0.99\linewidth]{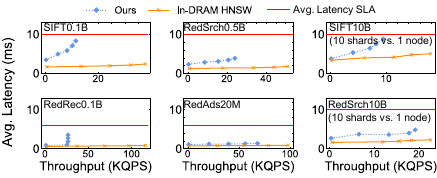}
	\centering
	\vspace{-3mm}
	\caption{\textit{Comparison with in-DRAM HNSW deployments.}}
	\label{fig:eva1-5}
	\vspace{-3mm}
\end{figure}

\noindent\textbf{Comparisons with in-DRAM deployments.} ~\autoref{fig:eva1-5} shows results. When HNSW fits in the same 96-core node (20M–0.5B datasets), \sysname achieves about 25–70\% of the throughput of in-DRAM HNSW deployments at 90\% recall, while always satisfying the average-latency SLAs. For the 10B-scale SIFT and RedSrch, the production HNSW  uses 10 shards with a total of 2.5 TB DRAM and 320 CPU cores (standard 32-core, 256-GB nodes). In contrast, \sysname reaches roughly 47-85\% of their throughput under the same SLA using a single 96-core machine with only 160–330 GB of DRAM. \sysname meets SLAs with modestly higher latency, but reduces CPU usage by about 3–4$\times$ and DRAM consumption by nearly an order of magnitude.

\subsection{Storage I/O Impact}
\label{sec:storage-eva}
 \begin{figure}[t]
	\centering
	\begin{subfigure}[t]{0.48\linewidth}
		\includegraphics[width=\linewidth]{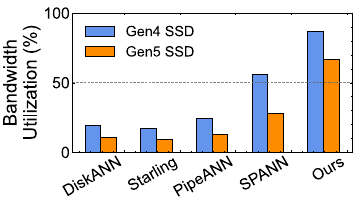}
		\caption{\textit{I/O Bandwidth utilization on PCIe-4.0 and PCIe-5.0 SSDs.}}
		\label{fig:evaS-1}
		\vspace{-3mm}
	\end{subfigure}
	\centering
	\begin{subfigure}[t]{0.48\linewidth}
		\includegraphics[width=\linewidth]{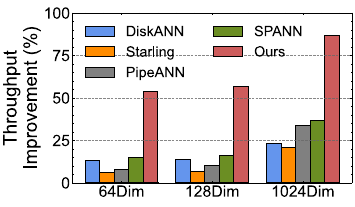}
		\caption{\textit{Improvements by upgrading SSDs from PCIe-4.0 to PCIe-5.0.}}
		\label{fig:evaS-2}
	\end{subfigure}
	\centering
	\caption{\textit{Impacts of different generations of NVMe SSDs.}}
	\vspace{-3mm}
	\label{fig:eva-Storage}
\end{figure}

\begin{figure}[t]
	\centering
	\includegraphics[width=0.92\linewidth]{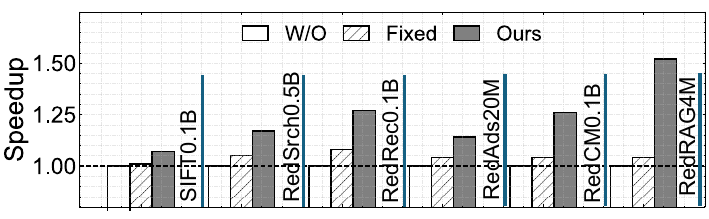}
	\centering
	\vspace{-2mm}
	\caption{\textit{Performance speedup by the pruning module.}}
	\label{fig:eva2-2}
		\vspace{-4mm}
\end{figure}
\noindent\textbf{Bandwidth utilization.} As shown in~\autoref{fig:evaS-1}, on RedSrch0.5B, we measure the bandwidth utilization of a 12-SSD array with both PCIe-4.0~\cite{PM9A3} and PCIe-5.0 drives~\cite{PM9D3A}. Graph-based systems utilize less than 20\% of SSD bandwidth, while clustering-based SPANN  also merely reaches about 55\% (Gen4). In contrast, \sysname pushes utilization to $\sim$85\% on Gen4 and $\sim$70\% on Gen5, showing that its batched, device-direct I/O path is much closer to the device limits.
\begin{figure}[t]
	\centering
	\includegraphics[width=0.99\linewidth]{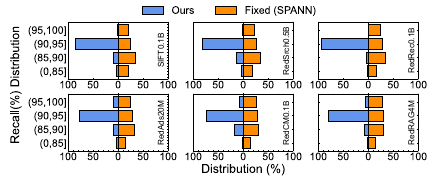}
	\centering
	\caption{\textit{Search quality improved by the pruning module.}}
	\label{fig:eva2-1}
	\vspace{-3mm}
\end{figure}

\noindent\textbf{Gains from hardware upgrades.} In~\autoref{fig:evaS-2}, we then report throughput gains when upgrading SSDs from PCIe 4.0 to 5.0 on 64-dim RedSrch, 128-dim RedAds, and 1024-dim RedRAG workloads. Graph-based systems (DiskANN, Starling, and PipeANN) improve by only 10-30\%, and SPANN by up to 40\%, indicating they are largely I/O-latency- or software-stack-bound. In contrast, \sysname gains about 55\% at 64- and 128-D and 87\% at 1024-D, demonstrating that its performance scales more efficiently with available bandwidth.

\subsection{Pruning Efficiency}
\label{sec:accuracy-eva}
For the public SIFT dataset, we randomly sample 1M vectors as the training set and use 10\% of them as queries; for production datasets, we extract 110K consecutive queries from online logs, use the first 100K as the training set, and reserve the last 10K as queries, consistent with our production scenario. By default, we set the number of iterations to 500 and the learning rate to 0.2.

\noindent\textbf{Performance gains.} Our pruning module consistently accelerates search across all datasets, as shown in~\autoref{fig:eva2-2}. Compared with the non-pruning baseline, it yields 1.1–1.6$\times$ throughput, and still provides 5–25\% higher throughput than the fixed pruning policy. The benefit is most pronounced on RedRAG4M because its workloads use small top-k values (i.e., 10–100), where the optimal search range (i.e., $nprobe$) tends to fluctuate more, leading to more redundant scans.

\noindent\textbf{Accuracy gains.} As shown in~\autoref{fig:eva2-1}, LLSP also leads to a more stable search quality under the same average recall. In contrast to the fixed policy, which fails to meet the target recall for over 40\% of queries, our method ensures that over 80\% of queries meet the target recall (i.e., 90\%), effectively reducing low-recall outliers across all six datasets.

\begin{table}[htbp]
\centering
\resizebox{\linewidth}{!}{
\begin{tabular}{ccc}
\toprule
\textbf{Dataset} & \textbf{Model} & \textbf{Feature Importance} \\
\midrule
\multirow{2}{*}{Red\textsf{Srch}} 
& Router & \(Query\):67.3\% $k$:32.7\% \\
& Pruning & \(Query\):34.3\% $k$:15.2\% \(Centroids\):50.5\% \\
\multirow{2}{*}{Red\textsf{RAG}} 
& Router & \(Query\):74.1\% $k$:25.9\% \\
& Pruning & \(Query\):48.3\% $k$:7.8\% \(Centroids\):43.9\% \\
\bottomrule
\end{tabular}}
\caption{Feature importance of leveling-learned pruning.}
\label{tab:feature_importance}
\end{table}

\noindent\textbf{Feature importance.} We further show the importance of different features in the leveling-learned pruning model in~\autoref{tab:feature_importance}. 
RedSrch0.5B and RedRAG4M are chosen as representatives of production datasets with different dimensions and top-$k$ ranges. RedSrch has a larger top-$k$ range (100–3,000) and smaller dimension (64), while RedRAG has a smaller top-$k$ range (10–100) and larger dimension (1024).
Results show that the query features (e.g., query coordinates and distances to centroids used during pruning) are important for both router and pruning models, indicating that the local distribution of queries is a key factor for determining the search range and pruning ratio.
Meanwhile, the importance of top-$k$ is higher for the router model than the pruning model, which is reasonable since the router determines the search range (i.e., $nprobe$) based on the target top-$k$, while the pruning model further prunes clusters within the smaller selected search range.

\subsection{Construction Evaluation}
\label{sec:build-eva}
 \begin{figure}[!t]
	\centering
	\begin{subfigure}[t]{0.48\linewidth}
		\includegraphics[width=\linewidth]{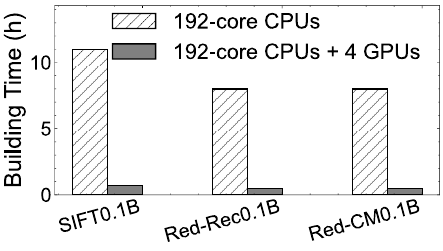}
		\caption{\textit{Comparison of single node.}}
		\label{fig:eva4-1}
	\end{subfigure}
	\centering
	\begin{subfigure}[t]{0.48\linewidth}
		\includegraphics[width=\linewidth]{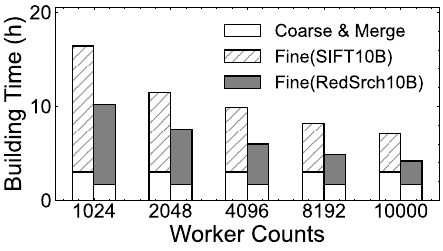}
		\caption{\textit{Speedup from elastic scaling.}}
		\label{fig:eva4-2}
	\end{subfigure}
	\centering
	\caption{\textit{Acceleration of GPUs and distribution.}}
	\label{fig:eva4}
	\vspace{-5mm}
\end{figure}
\noindent\textbf{GPU acceleration.} On a single 192-core node, CPU-only construction of 0.1B-scale indexes takes about 9–12 hours. Offloading the coarse clustering stage to 4 NVIDIA L20 GPUs reduces the total build time to within an hour, as shown in~\autoref{fig:eva4-1} (up to $\sim$10$\times$ speedup), transforming hours-long offline jobs into minute-level builds for high-freshness services (e.g., recommendation and advertising).

\noindent\textbf{Elastic scaling.} For 10B-scale datasets, using GPUs for coarse clustering and single-node merging for final indexes (i.e., Coarse \& Merge), we further parallelize the fine-grained balancing stage (i.e., Fine) across an elastic CPU cluster. Increasing the number of CPU workers from 1{,}024 to $10^4$ cores reduces end-to-end building time from over 16 hours to about 4--7 hours, as shown in~\autoref{fig:eva4-2}, enabling hours-level reconstruction of billion-scale indexes.

\subsection{Cost Efficiency}
\label{sec:cost-eva}
\begin{table}[ht]
	\centering
	\small
	\resizebox{0.99\linewidth}{!}{
		\begin{tabular}{lcccc}
			\toprule
			\multirow{2}{*}{\textbf{System}} &
			\textbf{DRAM} &
			\textbf{Gen5 SSD} &
			\textbf{Throughput} &
			\textbf{Storage Eff.} \\
			& \textbf{(GB, \$)} & \textbf{(GB, \$)} & \textbf{(KQPS)} & \textbf{(QPS/\$)} \\
			\midrule
			HNSW    & 123, 1K   & N/A         & 51  & 51 \\
			PipeANN & 8, 64     & 260, 52    & 0.8  & 7   \\
			SPANN   &  8, 64     &  162, 32     & 8.4 & 88  \\
			\textbf{Ours} &
			 8, 64     & 162, 32    & 24   & \textbf{250} \\
			\bottomrule
	\end{tabular}}
	\caption{Comparison of cost efficiency on RedSrch0.5B.}
	\label{tab:0.5B-cost}
	\vspace{-4mm}
\end{table}

\begin{table}[ht]
	\centering
	\small
	\resizebox{\linewidth}{!}{
		\begin{tabular}{lcccc}
			\toprule
			\multirow{2}{*}{\textbf{System}} 
			& \textbf{DRAM} 
			& \textbf{Gen5 SSD} 
			& \textbf{Throughput} 
			& \textbf{Storage Eff.} \\
			&\textbf{(TB, \$)} & \textbf{(TB, \$)} & \textbf{(KQPS)} &\textbf{(QPS/\$)} \\
			\midrule
			HNSW & 2.5, 20K & N/A & 23  & 1.2  \\
			\textbf{Ours} & 0.16, 1.3K & 3.2, 0.6K & 19  & \textbf{10 } \\
			\bottomrule
		\end{tabular}
	}
	\vspace{-2mm}
	\caption{Comparison of cost efficiency on RedSrch10B.}
	\label{tab:10B-cost}
	\vspace{-4mm}
\end{table}

\begin{table}[ht]
	\centering
	\small
	\resizebox{\linewidth}{!}{
		\begin{tabular}{lcccc}
			\toprule
			\multirow{2}{*}{\textbf{System}} 
			& \textbf{Cloud Price} 
			& \textbf{Build Time} 
			& \textbf{Offline Cost} 
			 \\
			&\textbf{(norm. price / hour)} & \textbf{(hour)}  &\textbf{(norm. cost)} \\
			\midrule
			HNSW & 1 (96 Core) & 1.5 & 1.5   \\
			\textbf{Ours} & 1.3 (96 Core + 4 GPU) & 1.3 & 1.7   \\
			\bottomrule
		\end{tabular}
	}
	\vspace{-2mm}
	\caption{Comparison of construction cost on RedSrch0.5B.}
	\label{tab:0.5B-construction}
\end{table}

\noindent\textbf{Cost efficiency of online serving.}
We compare storage cost and throughput on search workloads to show online serving costs. On RedSrch0.5B (\autoref{tab:0.5B-cost}), under production top-$k$ workloads, PipeANN sustains only 0.8 KQPS, yielding 7 QPS/\$, even lower than in-DRAM HNSW (51 QPS/\$). SPANN reaches 88 QPS/\$, while \sysname further improves to 250 QPS/\$ (5.4$\times$ over HNSW, 2.9$\times$ over SPANN). On RedSrch10B (\autoref{tab:10B-cost}), by replacing a 10-shard HNSW deployment using $\sim$2.5 TB DRAM with each single node using 32 cores and 256 GB DRAM, \sysname boosts cost efficiency from 1.2 to 10 QPS/\$ (8.3$\times$) by saving 90\% of DRAM usage  at an additional SSD cost of only \$0.6K.

\noindent\textbf{Cost efficiency of offline building.} In addition to compare our offline building time with the CPU-only SPANN, we also estimate the single-node CPU-GPU construction cost of \sysname and CPU-based HNSW we deployed previouly by comparing the normalized cloud price of CPU-only and CPU-GPU instances~\cite{EC2Price}. 
As shown in~\autoref{tab:0.5B-construction}, our GPU-accelerated construction achieves similar cost compared with CPU-only HNSW building, because of higher price of GPU instances but shorter build time. And as the GPU-based clustering implementation is developing~\cite{Flash-KMeans}, we expect the cost of clustering-based indexes' construction to further reduce, making it more cost-effective.

\section{Deployment}
\subsection{Deployment Status}
At the time of publication, \sysname has been rolling out as a unified ANN layer
in RedNote’s production. We use \sysname to support all services, and all-flash
servers are gradually replacing previous in-DRAM deployments across search,
recommendation, advertising, and other vector services in the coming months.
Currently, with only $\sim$40 all-flash servers-each with 0.7--1.1 TB DRAM and 12
NVMe SSDs-\sysname already sustains online workloads that previously required
about 35{,}000 CPU cores and $\sim$0.35 PB DRAM in in-DRAM deployments.
In
the offline construction phase, \sysname can leverage up to 10{,}000 CPU cores
from online clusters during low-traffic periods (e.g., late-night hours) to
accelerate large-scale builds. Looking forward, RedNote plans to migrate the
majority of in-DRAM deployments onto \sysname, with projected infrastructure cost
savings on the order of tens of millions of dollars per year.

\subsection{Operational Lessons and Future Work}
\textbf{Performance bottlenecks stem from local hotspots from queries and SSDs' die-level conflicts.} In most indexes, the overall deployment remains smooth and stable. When SSD
bandwidth and CPU utilization remain unsaturated, latency stays within SLAs and
throughput scales with additional search threads.  However, in the early stage
of trial operation, for a few recommendation indexes, adding CPU cores failed to
increase throughput, even though SSD bandwidth remained below 20\%.  Log replay shows that
transient query bursts can target the same nearest clusters and logical blocks,
causing internal chip conflicts~\cite{die-conf} and high-latency SSD accesses.
To mitigate this, we place a few redundant copies of the cluster lists on NVMe SSDs, which reduces conflicts and raises the throughput ceiling by 1.5–2$\times$, while incurring only a minor extra SSD space overhead.
\noindent\textbf{Memory bandwidth becomes the bottleneck before external I/O bandwidth is fully utilized.}
In our all-flash deployment, 12 Gen5 SSDs provide roughly 140 GB/s aggregate external I/O bandwidth, but only about 70\% can be utilized in practice. The bottleneck is the limited effective memory bandwidth of modern 12-channel DDR5 servers, typically around 300--350 GB/s, which must serve SSD-to-DRAM transfers, DRAM-to-CPU rereads, and nearest-centroid search. Consequently, adding more SSDs brings limited benefit once memory bandwidth is saturated, as observed in our early more SSDs equipments (e.g., 20 Gen5 SSDs experiments). This imbalance is likely to worsen as modern CPUs expose enough PCIe lanes (e.g., 128-160 lanes~\cite{9005CPUs}) to attach over 32 Gen5 SSDs, providing external bandwidth beyond what current memory systems can sustain. For the commercially available devices, the direct I/O-to-cache techniques such as Intel DDIO~\cite{DDIO} and AMD SDCI~\cite{SDCI} may mitigate this issue, but currently these methods only work for network devices and their SSD ecosystem remains immature. We think that extending such support to SSD-based systems might be a promising future direction. At the same time, recent studies~\cite{GuestANN,FlashANNS,TridentANN} have used GPUs to overcome the limitations of CPU compute power and host memory bandwidth, while also taking advantage of the increasingly abundant external I/O bandwidth provided by NVMe SSDs. We plan to explore their optimizations in our future deployments.

\noindent\textbf{Large top-$k$ can be more important than extremely high recall.}
In production search, recommendation, and advertising systems, first-stage ANNS is typically used as a candidate generator rather than the final decision maker. Since single-vector similarity only provides a compressed semantic signal, downstream stages usually apply post-filtering based on scalar attributes, or rescore the retrieved candidates using richer features and the original content, such as text, images, and videos ~\cite{TaobaoEBR,CLEAR,VBASE}. Therefore, in our workloads, retrieving a large candidate pool, e.g., top-$k$ from 100 to 1000 at around 90\% recall, is often more valuable than pursuing extremely high recall such as 98--99.9\% for small top-$k$ requests. The latter can substantially increase ANNS search cost but brings limited end-to-end business gain, because downstream stages mainly benefit from having enough valid and diverse candidates to filter and rescore, rather than from marginal improvements in the vector recall of a small candidate set. This suggests that future ANNS systems should be evaluated not only by recall@10~\cite{BigANN}, but also by performance under large top-$k$, which better match multi-stage pipelines.

\noindent\textbf{In-place updates remain challenging.}
Dynamic updates are important for freshness in vector data management, especially to keep up with the near-real-time and streaming training of embedding models. In a typical 0.1B-scale recommendation workload with hourly rebuilding and $\sim$25--30 KQPS of search, fully replacing periodic rebuilds with in-place updates would require roughly 25--30 KOPS of updates (i.e., insertions and deletions) while concurrently online search. This is beyond the update and search throughputs of recent state-of-the-art large-scale dynamic ANNS systems~\cite{SPFresh,OdinANN}. Therefore, supporting both high-rate updates and high-performance search remains a challenging direction.
Our current deployment adopts a hybrid design~\cite{Milvus,Manu,Neos,GPU-MS-ANNS}: the main SSD-resident index is periodically rebuilt, recent insertions are maintained in an auxiliary in-memory index e.g., (HNSW and IVF), and deletions are tracked by a tombstone bitmap. Queries search both indexes, merge candidates, and filter tombstoned vectors. This preserves freshness with modest complexity, but still incurs extra memory and rebuild costs.

\vspace{-2mm}
\section{Related Work}
\noindent\textbf{DRAM-SSD ANNS.} DRAM–SSD ANNS includes graph-based systems (e.g., DiskANN~\cite{DiskANN}, Starling~\cite{Starling}, PipeANN~\cite{PipeANN}) and clustering-based designs (e.g., SPANN~\cite{SPANN,SPFresh}). Though graph-based schemes are popular for issuing fewer I/Os than clustering-based search, our findings show that SSD latency and serial access can make them slower than clustering on modern SSDs with the near-DRAM bandwidth~\cite{KVell,ModernNVMe,NVMeArrays}, particularly for top-$k$ up to the thousands in production~\cite{TaobaoEBR,CLEAR,VBASE}. Hence, we integrate clustering methods with high-bandwidth NVMe SSD arrays.

\noindent\textbf{Pruning for ANNS.} Quake~\cite{Quake}, LAET~\cite{LAET}, Auncel~\cite{Auncel}, and ANSMET~\cite{ANSMET} study pruning and early stopping for in-DRAM ANNS, where the search can iteratively check intermediate results and terminate once convergence-like conditions are met. However, such fine-grained, stateful control conflicts with SSD-friendly batched I/Os, which prefer issuing large sequential reads without per-step feedback. Therefore, instead of these intermediate-result-based methods, we propose our pruning module trained on historical traces, which predicts search ranges without relying on intermediate states.

 \noindent\textbf{Acceleration of index building.} CAGRA~\cite{CAGRA}, RAFT~\cite{RAFT}, ParlayANN~\cite{ParlayANN}, and Faiss~\cite{Faiss} accelerate ANNS index construction and clustering using highly parallel graph builders and heterogeneous acceleration. We adapt several of their techniques (e.g., GPU-based $k$-means and parallel graph construction) in our offline index-building pipelines.

\vspace{-1mm}
\section{Conclusion}
Our experience at RedNote shows that, under industrial workloads with large top-$k$ queries, graph-based ANNS on SSDs suffers from long, latency-bound search paths and loses competitiveness. In contrast, pairing high-bandwidth NVMe arrays with the clustering-based search allows DRAM–SSD designs to approach in-DRAM performance (both throughput and latency) while greatly reducing hardware cost. \sysname demonstrates this path in practice, combining a custom storage stack, adaptive pruning, and fast elastic construction to deliver cost-effective, high-performance ANNS at scale.

\section*{Acknowledgment}
We thank all anonymous reviewers and our shepherd, Asaf Cidon, for their valuable feedback and helping us improve our paper significantly. 
We also thank engineers at Xiaohongshu Inc. for their efforts in deploying \sysname in production.
This work was conducted by Yuchen Huang and Baiteng Ma during their internship at RedNote Engine Architecture Department.
Erci Xu is the corresponding author.
Yuchen Huang and Chuliang Weng are supported by the National Natural Science Foundation of China (Grant No.62272171).

\bibliographystyle{plain}
\bibliography{HyperVec}

\end{document}